# Minimal Tools for Accurately Measuring the Coefficient of Kinetic Friction

Fathan Akbar, Nova Lailatul Rizkiyah, and Mikrajuddin Abdullah[a]

*Department of Physics, Bandung Institute of Technology*

*Jalan Ganesa 10 Bandung 40132, Indonesia*

(a)Email: mikrajuddin@gmail.com

Measuring the coefficient of kinetic friction is much more difficult than that of the static counterpart. Here we report a simple method using minimal tool to accurately measuring the coefficient of kinetic friction. We employed an inclined plane, the tool for measuring time, and a protractor. This method is much simpler and cheaper than other methods reported by some authors previously. The results are consistent with the date reported elsewhere.

**I. INTRODUCTION**

Measuring the coefficient of static friction is generally simple. For example, by using a smartphone and a chair to function as an inclines plane, Kapucu was able to determine the coefficient of static friction.[1] Kinsler and Kinzel measured the coefficient of static friction by employing very simple materials such as a meter stick, a protractor, and samples of the materials needed.[2] Dietz and Aguilar determined the coefficient of static friction by comparing the force needed to slip a block on a surface and a force needed to tip the block.[3] Other methods are available in internet resources for such a purpose.

To the contrary, the measurement of the coefficient of kinetic friction is much more difficult than that of static counterpart because the measurement must be conducted when the object is moving. The most commot most methods reported for measuring the coefficient of kinetic friction were applied by using a pulley system through which an object above the surface and a hanging object that attract the first object due to gravity is connected.[4] But the problem with this method is the omition the friction force experienced by the pulley so that the obtained friction force might be slightly different from the true value. Measuring the coefficient of kinetic friction by pulling the object on a surface using a spring balance generally results inconsistent values. The measured data will scatter when repeating the experiments.[5] Hue and Peachey measured the coefficient of kinetic friction by sliding down a block on an inclined surface and then allowed to free fall a certain height. Based on measurement how far the block displaced horizontally from the starting point of falling, the coefficient can be calculated.[5] However, in their experiment, the horizontal displacement of the block also varied in different trials, so that scattered data were still obtained even relatively small.

One automatic and simple method for measuring coefficient of kinetic friction either static or kinematic is using PASCO equipment. Indeed the equipment is a sensor for measuring the force experienced by object sliding on a surface. The change of force on time as the object is pulled can give rise the information of the static and kinetic forces and from both quantities, the coefficient of friction (static and kinetic) can be calculated easily.[6] This is is likely a friendly tool, but we must by it at a relatively high price. In addition, Lawlor reminded to be careful in using PASCO due to uncovering pre-sliding displacement which may give rise to different values of coefficients.[7]

An accurate method for measuring the coefficient of kinetic friction is using Timoshenko oscillator.[8] In this method, a solid (usually a long block) is placed onto two parallel cylinders separated by a distance of $L$. The two cylinders each rotate with the same angular speed ω, in opposite directions. The rotation speed must be large enough for the solid to slip on the cylinders so that the friction force is in the kinetic regime. Due to this rotation, the block center of mass oscillates with a period $T = \pi\sqrt{2L/\mu_k g}$ so that by measuring this period, the coefficient of kinetic friction can be determined. To obtain accurate data, the experiment set up must be fixed carefully and the selection of the best cylinder rotation speed should be determined carefully too.

In this paper, we report a simple method for measuring the coefficient of kinetic friction using minimal tools. This method can be applied elsewhere even in schools that are not provided by enough laboratory equipment. The key ingredient is by reorganizing the equation of motion under kinetic friction so that the new equation can be confronted with simple measurement.

## II. Derivation of Basic Equations

We place an object on an inclined plane so that it can slip down continuously at any speeds (**Fig. 1(A)**). The inclination angle must be high enough so that the component of gravitational force along the surface is larger than the kinetic frictional force directing upward along the surface. The total force experienced by the block is

$$F = W \sin\theta - f_k. \tag{1}$$

The kinetic friction force is given by $f_k = \mu_k N = \mu_k W \cos\theta$. Therefore, the acceleration of the object is $a = F/m$, or

$$a = g \sin\theta (1 - \mu_k \cot\theta) \tag{2}$$

We mark three points at positions $x=0$, $x=x_1$, and $x=x_2$ on the surface where two consecutive points are separated by the same distance, i.e.

$$x_1 - 0 = \Delta x \quad \text{and} \quad x_2 - x_1 = \Delta x \tag{3}$$

The block is released at any points to the left of $x=0$ so that it arrives $x=0$ with a speed $v_0$. We select the time reference so that $t=0$ is the time when the object just touches the position $x=0$. The speed of the object increases with time due to acceleration. The acceleration is assumed to be constant. The distance traveled by the object when arriving $x_1$ from the reference position ($x=0$) is

$$x_1 = \Delta x = v_0 t_1 + \frac{1}{2} a t_1^2 \tag{4}$$

Similarly, the distance traveled by the object when arriving $x_2$ from the reference position is

$$x_2 = 2\Delta x = v_0 t_2 + \frac{1}{2} a t_2^2 \qquad (5)$$

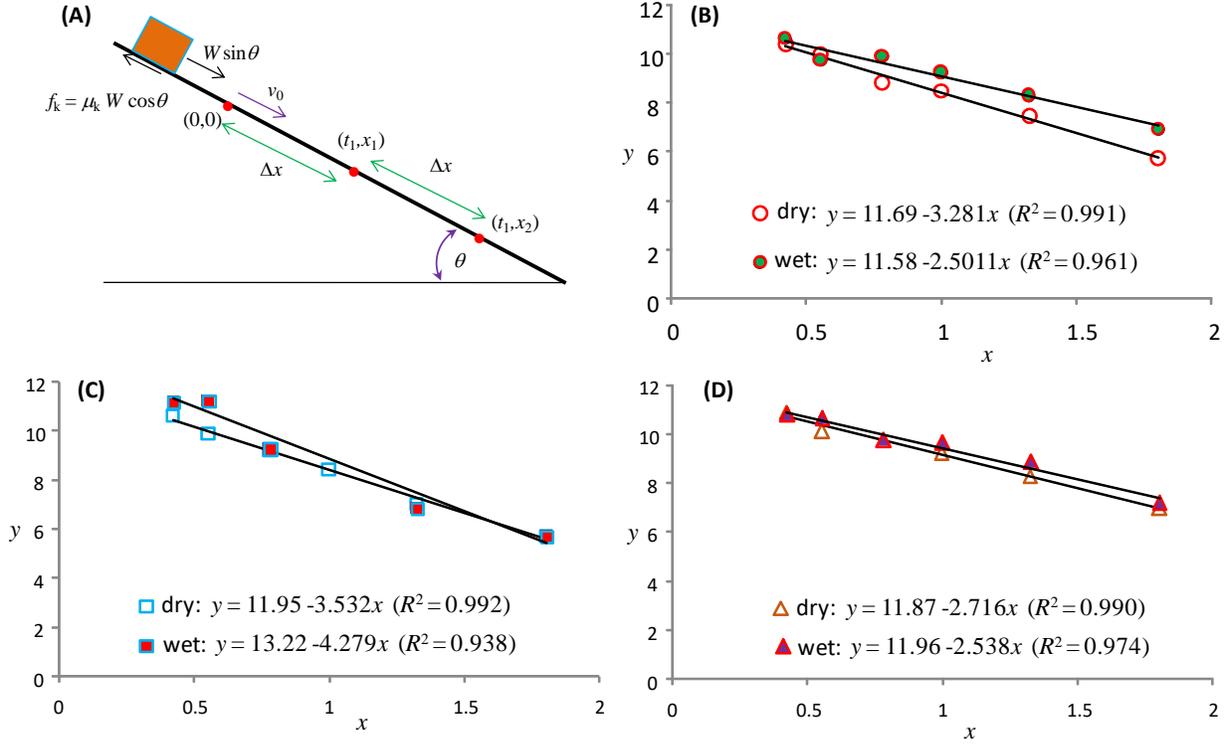

**Figure 1** (A) Schematic diagram for deriving equations. (B) – (D) Measurement results $y$ as function of $x$ ($y$ is defined in Eq. (12) and $x$ is defined in Eq. (13)). Symbols and measurement results in dry contact (open symbols) and wet contact (filled symbols). The linear equations are fitting for the data. From top to bottom: (B) aluminium on PVC, (C) wood on PVC, and (D) plastic tape on PVC.

Subtracting Eq. (5) with Eq. (4) one obtains

$$\Delta x = x_2 - x_1 = v_0(t_2 - t_1) + \frac{1}{2} a(t_2^2 - t_1^2)$$

$$= v_0(t_2 - t_1) + \frac{1}{2} a(t_2 + t_1)(t_2 - t_1) \qquad (6)$$

Now let us rewrite Eq. (4) as

$$\frac{\Delta x}{t_1} = v_0 + \frac{1}{2}at_1 \qquad (7)$$

and Eq. (6) as

$$\frac{\Delta x}{t_2 - t_1} = v_0 + \frac{1}{2}a(t_2 + t_1) \qquad (8)$$

Then we subtract Eq. (8) with Eq. (7) to obtain

$$\frac{\Delta x}{t_2 - t_1} - \frac{\Delta x}{t_1} = \frac{1}{2}a(t_2 + t_1) - \frac{1}{2}at_1$$

$$\frac{1}{t_2}\left(\frac{1}{t_2 - t_1} - \frac{1}{t_1}\right) = \frac{1}{2\Delta x}a \qquad (9)$$

Substituting the Eq. (2) into Eq. (9) we have

$$\frac{1}{t_2}\left(\frac{1}{t_2 - t_1} - \frac{1}{t_1}\right) = \frac{1}{2\Delta x}g(\sin\theta - \mu_k \cos\theta)$$

or

$$\frac{1}{t_2}\left(\frac{1}{t_2 - t_1} - \frac{1}{t_1}\right)\frac{1}{\sin\theta} = \frac{g}{2\Delta x} - \frac{g\mu_k}{2\Delta x}\cot\theta \qquad (10)$$

Eq. (10) is a common linear equation,

$$y = a + bx \qquad (11)$$

with

$$y = \frac{1}{t_2}\left(\frac{1}{t_2 - t_1} - \frac{1}{t_1}\right)\frac{1}{\sin\theta} \qquad (12)$$

$$x = \cot\theta \qquad (13)$$

$$a = \frac{g}{2\Delta x} \tag{14}$$

$$b = -\frac{g\mu_k}{2\Delta x} \tag{15}$$

For determining the coefficient of kinetic friction we only need to measure $t_1$, $t_2$ and $\theta$. We measure sets of $(t_1,t_2,\theta)$ at several $\theta$s and put the result in a graph of $y$ as a function of $x$. We can freely choose $\theta$ as long as the object can continuously slide down the plane. The obtained data are then fitted with a straight line $y = a + bx$. We can fit using a simple instruction in Excel, **Add Trendline**. Based on Eqs. (14) and (15), the coefficient of kinetic friction can be simply calculated as

$$\mu_k = -\frac{b}{a} \tag{16}$$

It seems that the measuring of the coefficient of kinetic friction is very simple and need only very minimum tools.

### III. Experiment

The crucial tool need for this experiment is a stopwatc for measuring the time used of the object to move from $x = 0$ to $x = x_1$ and from $x = 0$ to $x = x_2$. Other tools as easily obtained such as an inclined plane, various sliding objects, and a protractor. We might use the stopwatch provided by the smartphone if it is possible to get accurate data. This measurement can be accurate if the separation of two nearest points is large enough. However, if the separation is not so large, the measurement should not be done manually.

In this work we determined the times using light sensors placed at $x = 0$, $x_1$, and $x_2$. We connected the light sensors with an Arduino UNO where the voltages at the sensor were used as inputs at three analog pins of the Arduino. We recorded the time using **millis()** instruction so that the accuracy of recorded time is in milliseconds.

We used a roof gutter made of polyvinyl chloride (PVC) as the inclined plane. We used aluminium, wood, and tape plastic blocks, either in dry contact and wet contact (wetted with water) as sliding objects.

## IV. Results and Discussion

**Figure 1** (B)-(D) are the measured data for (B) aluminium on PVC, (C) wood on PVC, and (D) plastic tape on PVC. All data have been well fitted with a linear equation with $R^2 > 0.938$ to indicated the linear change is very acceptable. By using the data from fitting processes and Eq. (16) we obtain the coefficient of kinetic friction as listed in **Table 1**.

**Table 1** The measured coefficients of kinetic friction

| Contacting surface | $\mu_k$ at dry contact | $\mu_k$ at wet contact |
|---|---|---|
| PVC-aluminium | 0.281 | 0.216 |
| PVC-wood | 0.296 | 0.324 |
| PVC-plastic tape | 0.229 | 0.212 |

We can compare the results with data reported elsewhere. Murase reported that the coefficient of kinetic friction of PVC-wood at dry contact was approximately 0.3,[9] very consistent with our measurement. Using Timoshenko oscillator method, Henaff et al[8] measured the coefficient of kinetic friction between construction wood and rotating PVC cylinder at room temperature was 0.31±0.03. In our experiment, we used the construction wood which is very comparable to the result of Henaff et al.[8] The coefficient of kinetic friction dry PVC-metal around 0.25,[10] very close to our result for PVC-aluminium in dry condition of 0.281.

In general, the coefficient of friction decreases when the surface is wetted. This change is observed at the contacts between PVC-aluminium and PVC-plastic tape. But different behavior was observed at the contact between PVC-wood where the coefficient increases when the surface is wetted. Koubek and Dedicova showed that the coefficient of friction between wood and metal increases with increasing the wood moisture.[11] McKenzie and Karpovich investigated the coefficient of friction between steel and various woods and arrived at the general conclusion that the wet condition produces a larger coefficient than the dry condition.[12]

The above discussion clearly showed the proposed method is very useful and very easy to determine the coefficient of kinetic friction which so far has been difficult to be measured accurately. This method can be applied in all schools throughout the world and can become an interesting physics learning tool.

**VI. Conclusion**

We have shown that the method proposed here is very simple for measuring the coefficient of kinetic friction using minimal tools. The results are consistent with data reported elsewhere. This method is much simpler and cheaper than other methods reported by some authors previously. This method can be used elsewhere around the world without the need for tool investment.

**Acknowledgement**

This work was supported by Riset ITB 2019 from Bandung Institute of Technology.